\documentclass[a4paper]{jpconf}
\usepackage{amssymb}
\usepackage{graphicx}

\newcommand{\be}{\begin{equation}}
\newcommand{\ee}{\end{equation}}
\newcommand{\bra}{\langle}
\newcommand{\ket}{\rangle}
\newcommand{\bea}{\begin{eqnarray}}
\newcommand{\eea}{\end{eqnarray}}
\newcommand{\dis}{\displaystyle}

\begin{document}
\title{Bayesian estimation of realized stochastic volatility model by Hybrid Monte Carlo algorithm}

\author{Tetsuya Takaishi}

\address{Hiroshima University of Economics, Hiroshima 731-0192, JAPAN}

\ead{tt-taka@hue.ac.jp}

\begin{abstract}
The hybrid Monte Carlo algorithm (HMCA) is applied for Bayesian parameter estimation of 
the realized stochastic volatility (RSV) model.
Using the 2nd order minimum norm integrator (2MNI) for the molecular dynamics (MD) simulation in the HMCA,
we find that the 2MNI is more efficient than the conventional leapfrog integrator. 
We also find that the autocorrelation time of the volatility variables sampled by the HMCA 
is very short.
Thus it is concluded that the HMCA with the 2MNI is an efficient algorithm for
parameter estimations of the RSV model.
\end{abstract}

\vspace{-5mm}
\section{Introduction}
In empirical finance, volatility of asset returns plays an important role to manage financial risk.
Since volatility is not directly observed in financial markets, one needs to estimate volatility from
observed quantities in the markets.
A typical way to estimate volatility is to use parametric models which mimic properties of asset returns.
One of such models is the stochastic volatility (SV) model\cite{Taylor} which allows the volatility to be 
a stochastic process. 
Recently an SV-type model that further utilizes daily realized volatilities (RV) \cite{RV1,RV2} 
was proposed\cite{RSV}. This model is called the RSV model.
An advantage of the RSV model is that it uses the RV as additional information to 
the model and thus can estimate the daily volatility  more accurately.

To estimate parameters of SV-type models
one usually use the Markov Chain Monte Carlo (MCMC) method based on the Bayesian approach\cite{SV1,SV2,Multimove}.
In the MCMC method of the SV model 
the most time-consuming part is the volatility update.
The number of the volatility variables to be updated increases  with the data size of time series.
Generally a local update algorithm such as the Metropolis algorithm is not efficient for the update of volatility variables.

In this study we use  an alternative MCMC method, the HMCA\cite{HMC} to
estimate the parameters of the RSV model.
The main advantage of the HMCA is that
it is a global update algorithm 
that can update volatility variables simultaneously.
The HMCA has been tested for the SV model\cite{Takaishi1,Takaishi2,Takaishi3} and
it is found that the HMCA can de-correlate MCMC samples of volatility variables faster than 
the Metropolis algorithm. 
In this study we investigate the performance of the HMCA for the RSV model.
We also examine the efficiency of an improved integrator 
called the 2nd order minimum norm integrator (2MNI) 
used for the MD simulation in the HMCA.

\section{Realized Stochastic Volatility Model}

The basic realized stochastic volatility (RSV) model by Takahashi {\it et al.}\cite{RSV} is  given by
\bea 
y_t & = &\exp(h_t/2)\epsilon_t, \hspace{5mm} \epsilon_t \sim N(0,1), \\
\ln RV_t & = & \xi + h_t + u_t, \hspace{5mm} u_t \sim N(0,\sigma_u^2), \\
h_{t+1}& = & \mu +\phi(h_t-\mu)+\eta_t,\hspace{5mm}  \eta_t \sim N(0,\sigma_{\eta}^2), 
\eea
where $y_t$ and $RV_t$ are a daily return and a daily realized volatility  at time $t$ respectively, 
and $h_t$ is a latent volatility defined by $\ln \sigma_t^2$.
This model contains 5 parameters (${\bf \theta }=\phi,\mu,\xi,\sigma_{\eta}^2,\sigma_u^2$)
that have to be determined so that the model matches the underlying returns and realized volatilities. 
We apply the Bayesian inference for parameter estimations
and perform the Bayesian inference by the MCMC method. 
For the MCMC simulation of volatility variables 
we use the HMCA. 
On the other hand, we update model parameters 
by the standard MCMC method previously 
used in parameter estimations of the SV model\cite{SV1,SV2,Multimove,Takaishi1}. 

\section{Hybrid Monte Carlo Algorithm}
Originally the HMCA was invented and developed for the MCMC simulations of
the lattice Quantum Chromo Dynamics (QCD) calculations\cite{HMC}
where non-linear interactions  between gauge variables are major obstacles 
to perform the MCMC method\cite{LatticeBook}.
The basic idea of the HMCA 
is a combination of the MD simulation and the Metropolis accept/reject test. 
Candidates for next volatility variables in Markov chain 
are obtained by solving 
the Hamilton's equations of motion (HEOM) in fictitious time $\tau$,
$ \dis
\frac{dh_i}{d\tau}=\frac{\partial H}{\partial p_i},
\hspace{2mm}
\frac{dp_i}{d\tau}=-\frac{\partial H}{\partial h_i},
$
where $p_i$ is a conjugate momentum to $h_i$.
The Hamiltonian $H$ is defined by 
$H(p,h)=\frac12 \sum_i^n p_i^2 - \ln f(h,\theta)$,
where $f(h,\theta)$ is the conditional posterior density of the RSV model\cite{RSV}.
Since in general HEOM
can not be solved analytically, we integrate them numerically through the MD simulation.
The conventional integrator for the MD simulation in the HMCA 
is the 2nd order leapfrog integrator (2LFI)\cite{HMC} given by
\bea
& & h_i(\tau +\delta \tau/2) =  h_i(\tau)+\frac{\delta \tau}{2}p_i(\tau),  \hspace{5mm}
p_i(\tau+\delta \tau)  =  p_i(\tau)- \delta \tau \frac{\partial H}{\partial h_i}, \\ \nonumber
& & h_i(\tau +\delta \tau) =  h_i(\tau+\delta \tau/2)+\frac{\delta \tau}{2}p_i(\tau+\delta \tau), 
\label{eq:2LF}
\eea
where $\delta \tau$ stands for the step size in the MD simulation.
In the operator representation the 2LFI can be expressed as 
$\dis e^{\lambda\delta\tau T/2}e^{\delta\tau V} e^{\lambda\delta\tau T/2}$,
where $T$ and $V$ are the linear operators corresponding to the kinetic term and potential term of 
the Hamiltonian respectively\cite{Forest,Yoshida}
Eqs.(\ref{eq:2LF}) are repeatedly applied and, volatility and momentum variables are  
integrated  up to a certain time $l$ called the total integration length.  
The numerical integration introduces the integration errors that violate the conservation of the Hamiltonian.
Let $h^{\prime}$ and $p^{\prime}$ be  volatilities and momenta after integration.
Then we define the difference of the Hamiltonian between after and before integration as
$\Delta H= H(p^{\prime},h^{\prime})-H(p,h)$.
Finally  new candidates for  volatilities $h_t^{\prime}$ 
are accepted according to
the Metropolis probability, $\min\{1,\exp(-\Delta H)\}$. 

The magnitude of $\Delta H$ depends on the integrator we choose.
It is possible to use higher order integrators (HOI) that can decrease the magnitude of $\Delta H$ at the same step size
and increase the Metropolis acceptance.
On the other hand HOI are usually computationally expensive. 
Therefore the total performance of HOI is not known a priori 
and depends on the model we use\cite{Takaishi1,Takaishi2}. 
In Ref.\cite{Takaishi3} it is shown that an improved integrator called 
the 2MNI{\it et al.}\cite{Omelyan} 
is superior to the 2LFI for the lattice QCD simulations.
The 2MNI is expressed as  $ \dis e^{\lambda\delta\tau T}e^{\delta\tau V/2}  
e^{(1-2\lambda)\delta\tau T}e^{\delta\tau V/2}e^{\lambda\delta\tau T}$
where $\lambda$ is a tunable parameter. 
$\lambda$ is set to 0.193183327. This value is approximately the optimum 
value that minimizes the error function of 2MNI\cite{Omelyan}. 
In this study we also use the 2MNI for the HMCA  
and examine its efficiency for the HMCA of the RSV model.

\section{Simulation Study}
We generated artificially 4000 data with RSV parameters 
$(\phi,\mu,\xi,\sigma_{\eta}^2,\sigma_u^2)=(0.93,-1.0,0.3,0.1,0.2)$
and performed the Bayesian estimation of the RSV model for this artificial data.

Fig. 1 shows the root mean square (RMS) of $\Delta H$ as a function of $\delta \tau$.
Here the total integration length is set to 2.
At small $\delta \tau$ the RMS of $\Delta H$ for both integrators is proportional 
to $\delta \tau^2$ as theoretically expected\cite{Karsch,HOHMC1}.
As $\delta \tau$ increases higher order terms contribute the RMS of $\Delta H$ and 
the RMS of $\Delta H$ increases more rapidly with $\delta \tau$.
The efficiency of the integrators can be measured by the efficiency function 
defined by $P(\delta\tau)\delta\tau$ where $P(\delta\tau)$ stands for the acceptance at $\delta\tau$.
Fig.2 shows the efficiency function as a function of  $P(\delta\tau)$.
It is found that the efficiency function takes a maximum at a certain optimum 
acceptance $P_{opt}$.
For the 2LFI, $P_{opt}$ is found to be around 0.65 which is consistent
with the theoretical value 0.61\cite{HOHMC1} for the 2nd order integrators.
On the other hand for the 2MNI,  $P_{opt}$ is around $0.8\sim 0.9$ which is 
rather close to the optimum value for the higher order integrators\cite{Takaishi1}.
This can be explained by that higher order error terms of the 2MNI dominate the RMS of $\Delta H$   
already at lower $\bra \Delta H^2\ket ^{1/2}$.
From fig.2  we see that the efficiency function of the 2MNI at the optimum is about 5 times
bigger than that  of the 2LFI.
Since the computational cost of the 2MNI is 2 times bigger than that of the 2LFI\cite{Takaishi3},
in total the 2MNI is about 2.5 times more efficient than the 2LFI.

Table 1 shows the results obtained by the MCMC simulations. 
Here the HMCA with the 2MNI is performed with $\delta \tau=0.222$ which corresponds to the optimum acceptance.
We find that the results obtained by the MCMC simulations
are consistent with the input values,
which means that our MCMC algorithm worked correctly.
The autocorrelation time (ACT), $2\tau_{int}$ is defined by 
$ 2\tau_{int} =1+2\sum_{t=1}^{\infty} ACF(t)$ where $ACF(t)$ stands for  the autocorrelation function.
In table 1 we only show the results of $h_{10}$ as a representative one for the volatility variables. 
$2\tau_{int}$ of  $h_{10}$ is  around $21$ which is very small compared to about 200 of the Metropolis algorithm\cite{Takaishi2}.

\begin{figure}[t]
\begin{minipage}{0.5\hsize}
\begin{center}
\includegraphics[height=5cm]{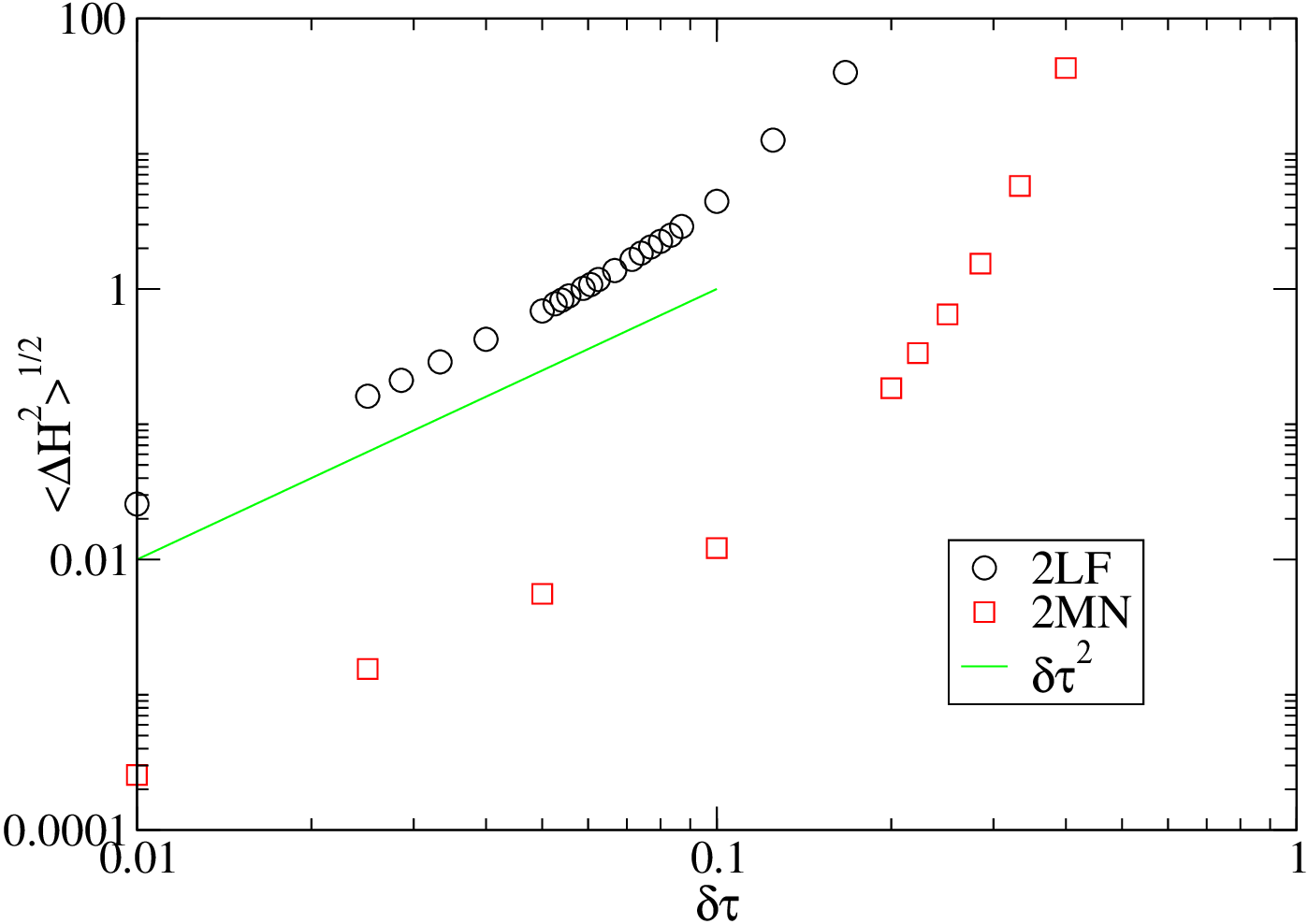}
\caption{ $\bra \Delta H^2\ket ^{1/2}$ as a function of $\delta \tau$.
The solid line shows a line proportional to $\delta \tau^2$.
}
\label{fig:dH}
\end{center}
\end{minipage}
\hspace{3mm}
\begin{minipage}{0.5\hsize}
\begin{center}
\includegraphics[height=5cm,keepaspectratio=true]{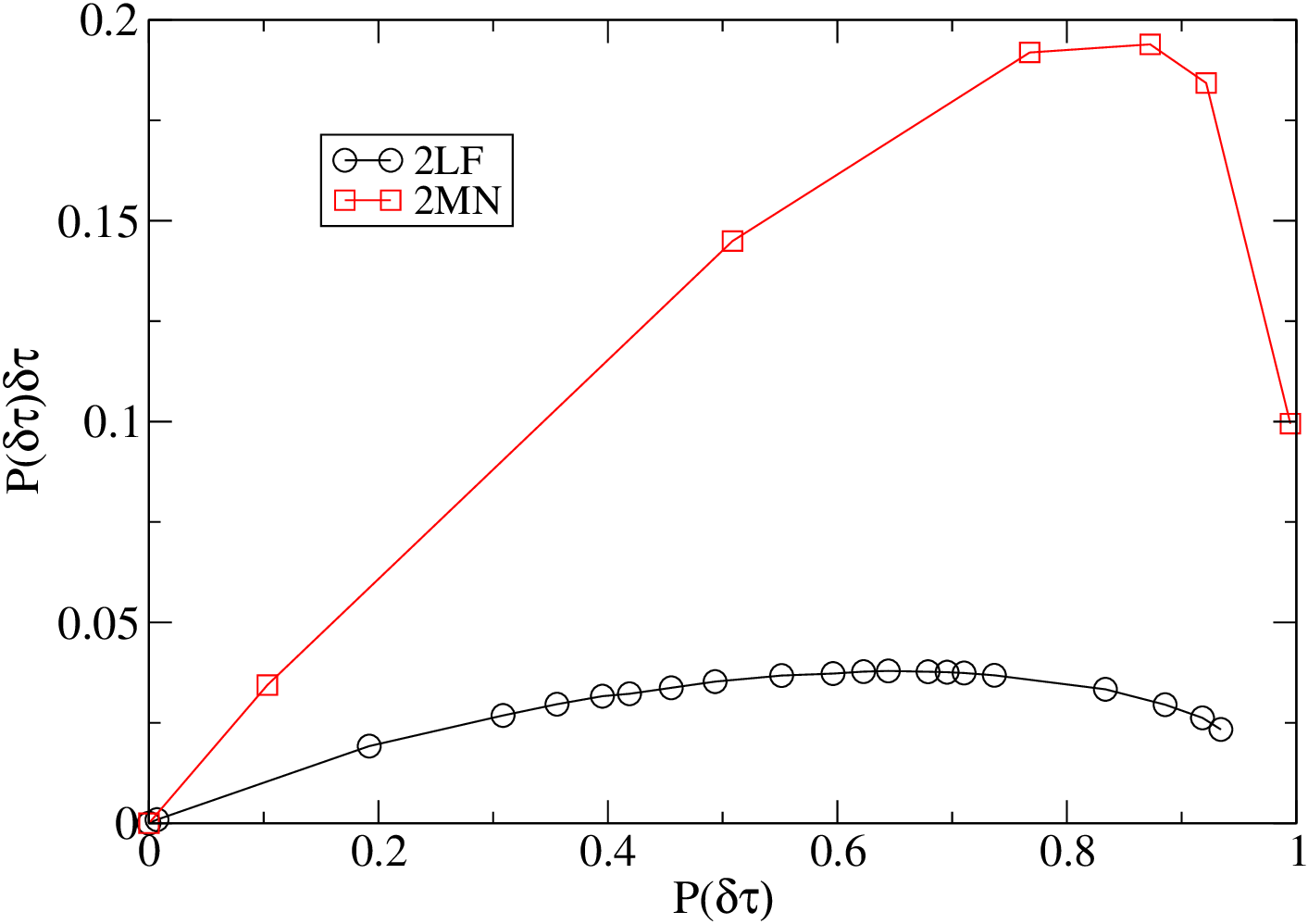}
\caption{The efficiency function $P(\delta \tau)\delta \tau$ versus $P(\delta \tau)$.
}
\label{fig:Acc}
\end{center}
\end{minipage}
\end{figure}

\begin{table}[t]
\vspace{-2mm}
\centering
\caption{Results obtained by MCMC simulations. S.D. stands for standard deviation.
}
\label{tb:data1}
\begin{tabular}{c|cccccc}
\hline 
               & $\phi$ & $\mu$ & $\xi$ & $\sigma_{\eta}^2$ & $\sigma_u^2$ &  $h_{10}$ \\
\hline
input          & 0.93   & -1.0  & 0.3  & 0.1   & 0.2   &     \\
average        & 0.926  & -0.97 & 0.31 & 0.097 & 0.203 & -9.77 \\
S.D.           & 0.007  & 0.10  & 0.03 & 0.006 & 0.010 & 0.23 \\
$2\tau_{int}$  & 8.8(20) & 96(40) & 1130(480) & 64(18) & 38(10) & 21(6)  \\
\hline
\end{tabular}
\vspace{-2mm}
\end{table}

\section{Empirical Study}
As an empirical study we also performed the Bayesian inference of the RSV model
for the stock price data of Panasonic Co. sampled from 3 July 2006 to 30 Dec. 2009
and estimated the model parameters of the RSV model. 
The RV is constructed by using 
1min-sampling high-frequency returns.
We use the 2MNI for the HMCA and
set the step size to 0.1 and the total integration length to 1.
We discarded the first 5000 MCMC samples and then accumulated
50000 samples for analysis.

Table 2 shows the results estimated from the MCMC simulation.
The ACT of $h_{10}$ is found to be short, around 37.
Thus MCMC samples of volatility variables are well de-correlated by the HMCA.
The parameter $\xi$ corresponds to the bias correction of the RV\cite{RSV}.
The raw RV is distorted by bias such as microstructure noise and non-trading effects.
To correct such bias Hansen and Lunde\cite{HL} introduced 
an adjustment factor $c$ which rescales the average of the RV to the variance of 
the daily returns. Using $c$, the raw $RV$ is corrected to $cRV$.
If $\xi$ explains the same bias as the adjustment factor $c$
the relation $\xi =-\log(c)$ should hold.
For the RV with 1min sampling frequency we obtain $c=1.244$, thus $-\log(c)=-0.2183$.
On the other hand we obtain $\xi=-0.101$ which is considerably different from $-\log(c)$.
Therefore this indicates that $\xi$ may also include other type of bias in the RV. 

\begin{table}[th]
\vspace{-2mm}
\centering
\caption{Results obtained by MCMC simulations for Panasonic Co.  }\label{tb:data2}
\begin{tabular}{c|cccccc}
\hline 
               & $\phi$ & $\mu$ & $\xi$ & $\sigma_{\eta}^2$ & $\sigma_u^2$ &  $h_{10}$ \\
\hline
average        & 0.958  & -7.63 & -0.101 & 0.0440 & 0.0176 & -7.21 \\
S.D.           & 0.013  & 0.14  & 0.049 & 0.0036 & 0.0031 & 0.12 \\
$2\tau_{int}$  & 39(6) & 24(0) & 178(42) & 65(11) & 112(19) & 37(11)  \\
\hline 
\end{tabular}
\vspace{-2mm}
\end{table}

\section{Conclusion}
We have performed the Bayesian estimation of the RSV model 
by the HMCA.
We find that the HMCA with the 2MNI is more efficient than 
the conventional 2LFI. 
We also find that the ACT of the volatility variables by the HMCA 
is short, which means that the HMCA can sample well de-correlated volatility variables.
Finally we conclude that 
the HMCA with the 2MN integrator is efficient for 
the Bayesian parameter estimation of the RSV model.

\section*{Acknowledgement}
Numerical calculations in this work were carried out at the
Yukawa Institute Computer Facility
and the facilities of the Institute of Statistical Mathematics.
This work was supported by JSPS KAKENHI Grant Number 25330047.

\end{document}